\documentclass[11pt]{article}

\usepackage[final]{acl}

\usepackage{times}
\usepackage{latexsym}
\usepackage{booktabs}
\usepackage{multirow}
\usepackage{soul}
\usepackage{graphicx}    
\usepackage{amsmath}     
\usepackage{amssymb}     
\usepackage{xcolor}      
\usepackage{enumitem}
\usepackage{makecell}
\usepackage{xcolor}
\usepackage{graphicx}
\usepackage{sourcesanspro}
\usepackage[most]{tcolorbox}
\usepackage{wrapfig}
\usepackage{needspace}

\usepackage{amsthm}
\newtheorem{definition}{Definition}
\definecolor{promptpink}{RGB}{231, 142, 144}

\newtcolorbox{promptbox}[1][Prompt]{
  enhanced,
  breakable,
  colback=promptpink!5,
  colframe=promptpink!5,
  boxrule=0pt,
  arc=2pt,
  leftrule=3pt,
  colframe=promptpink!80!black,
  left=8pt, right=6pt, top=6pt, bottom=6pt,
  fontupper=\fontfamily{SourceSansPro-LF}\selectfont\footnotesize,
}

\usepackage[T1]{fontenc}

\usepackage[utf8]{inputenc}
\usepackage{pifont}
\usepackage{microtype}
\usepackage{makecell}
\usepackage{inconsolata}

\usepackage{graphicx}
\usepackage{subcaption}
\title{Beyond a Single Story: Meta-Reviewing Sparse and Incomplete \\User-generated Contents for Recommendation}

\author{
\textbf{Hongren Wang\textsuperscript{1,\ding{170}}},
\textbf{Tianjun Wei\textsuperscript{2}}\thanks{Corresponding author.},
 \textbf{Yingpeng Du},
 \textbf{Jie Zhang},
 \textbf{Yin-Leng Theng}
\\
 Nanyang Technological University, Singapore
\\
\textsuperscript{1}\href{mailto:first.author@ntu.edu.sg}{\texttt{hongren001@e.ntu.edu.sg}} \quad
\textsuperscript{2}\href{mailto:second.author@ntu.edu.sg}{\texttt{tjwei2-c@my.cityu.edu.hk}}
 }

\begin{document}
\maketitle
{\renewcommand{\thefootnote}{\ding{170}}\footnotetext{Collaborative Initiative, Interdisciplinary Graduate Programme, Nanyang Technological University, Singapore; College of Computing and Data Science, Nanyang Technological University, Singapore.}}

\begin{abstract}
Data sparsity remains a long-standing challenge in recommender systems, and it becomes more severe for methods relying on user-generated content (UGC) such as textual reviews, which capture fine-grained preferences but require more user efforts to produce. As a result, UGC exhibits (1) \textbf{missing reviews}, where interactions lack any review, and (2) \textbf{incomplete reviews}, where available reviews cover only a subset of relevant attributes. Existing approaches often overlook these UGC-specific issues, leading to degraded accuracy. Motivated by \textbf{meta-review} in academic peer review, we propose \textsc{Mosaic} (\textbf{M}eta-review \textbf{O}n \textbf{S}parse \textbf{A}nd \textbf{I}ncomplete user-generated \textbf{C}ontent), which constructs a meta-review for each target user by aggregating attribute-sentiment evidence from neighbor users' reviews.  A multi-gate mixture-of-experts (MMoE) architecture jointly optimizes rating prediction and meta-review attribute-sentiment prediction, while an attention module personalizes the aggregated meta-review signals to each target user, yielding both refined rating predictions and attribute-level explanations. Experiments on four real-world datasets demonstrate that \textsc{Mosaic} consistently outperforms state-of-the-art baselines in both recommendation accuracy and explanation quality, mitigating UGC sparsity and incompleteness while delivering consistent gains for users with \textbf{limited interaction history}.
{\renewcommand\thefootnote{\textdagger}\footnote{Code is available on GitHub: \url{https://github.com/wanghrrrr1207/MOSAIC-Recommendation}.}}

\end{abstract}


\maketitle

\section{Introduction}
The rapidly expanding volume of digital information and user data has made recommender systems (RSs) an indispensable tool for personalizing user experience and driving business growth across various Internet and web platforms~\cite{li2024recent, gheewala2025depth}. Yet a foundational challenge has persisted at the core of these systems: data sparsity \cite{gunathilaka2025addressing, gheewala2025depth}. Most prior work alleviates the \textbf{sparsity of behavioral feedback}, such as clicks and ratings~\cite{cheng2023multi, lin2024inverse, liu2023multi}, but such signals capture only coarse-grained preferences and 
miss user reasoning and context. User-generated content (UGC), such as reviews, is semantically rich and complements numerical ratings~\cite{isinkaye2015recommendation}. However, since writing reviews requires considerable user effort, UGC itself is even sparser than behavioral feedback, making the 
\textbf{sparsity of UGC a more severe and critical problem}.


\begin{figure}[t]
\vspace{2em}
  \centering
    \includegraphics[width=\linewidth]{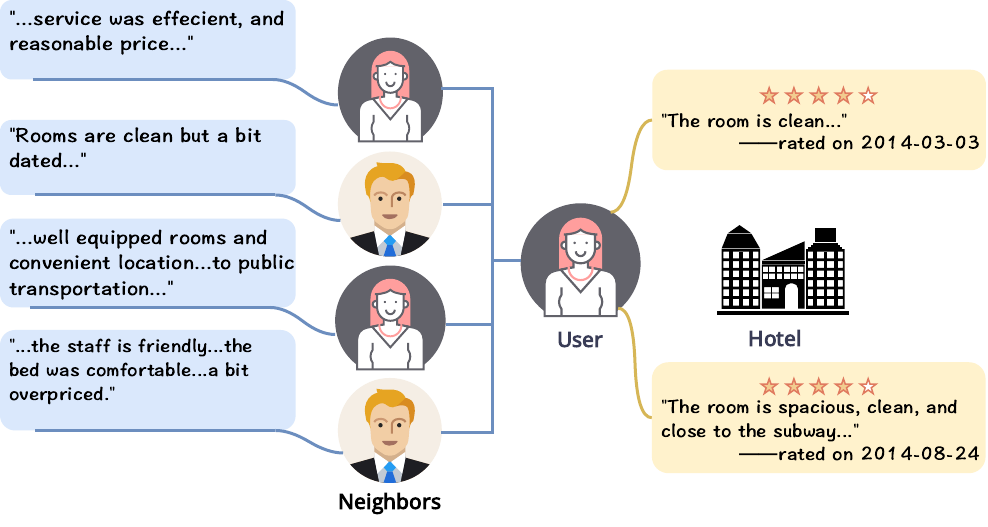}
    \caption{A \textit{TripAdvisor} example illustrating review incompleteness: the same user mentions different attributes in different reviews of the same hotel. Aggregating neighbor reviews as a meta-review compensates for the missing attributes.}
    \label{fig:user_review}
\end{figure}

To combat this, advanced Natural Language Processing (NLP) techniques have been applied to extract fine-grained information from reviews \cite{wei2023expgcn,Liu_2025,liao2025aspect}. Early aspect-based methods~\cite{dong2020asymmetrical, shuai2022review} identified 
opinions linked to product attributes but often struggled with 
noisy and unstructured text. Recent work~\cite{Zhang_2024} leverages Large Language Models (LLMs) to extract a more precise extraction of item-related characteristics. Going further, LLM-based recommenders such as EXP3RT~\cite{EXP3RTSIGIR25} treat LLMs as \emph{review interpreters}, distilling reasoning from a teacher LLM to extract preferences from raw reviews for rating prediction. However, these methods fundamentally process information 
\emph{available within each user's own reviews}, leaving user-item pairs without review content weakly supervised and the alleviation of UGC sparsity bounded by what individual users have written.


Beyond these limitations, a more subtle and granular form of data sparsity persists: the \textbf{inherent incompleteness of individual reviews}. As shown in Figure \ref{fig:user_review}, users rarely provide exhaustive reviews that fully encapsulate their complete set of preferences, leaving attribute-level gaps in their preference profiles. For instance, a user might praise a tablet's "\textit{high-resolution screen}" but omit the \textit{``battery life''} that also matters to them. \textbf{Relying on a single user's reviews is often insufficient to fully capture their diverse preferences}. Existing NLP or LLM-based extraction methods cannot compensate for this absence, since they can only identify attributes explicitly mentioned in the text. While LLM-based recommenders such as LLMRec~\cite{wei2024wsdm} can generate auxiliary content to enrich sparse side information, such
content remains susceptible to LLM hallucination and lacks verifiable grounding in real user feedback. This subtle yet critical form of sparsity significantly limits the ability of RSs to construct holistic user representations, hence motivating the development of strategies that enrich individual reviews with complementary information grounded in real user evidence.

In this work, we propose \textsc{Mosaic} 
(\textbf{M}eta-review \textbf{O}n \textbf{S}parse \textbf{A}nd \textbf{I}ncomplete user-generated \textbf{C}ontent), which introduces the idea of \textit{meta-review} from the academic peer review process~\cite{bhatia2020metagen} to address this fine-grained sparsity. \textsc{Mosaic} hypothesizes that aggregating reviews from neighbor users with similar preferences yields a more comprehensive view than any single review. Unlike prior work that aggregates neighborhood information at the review or instance level\cite{shuai2022review}, our meta-review construction operates at the \textbf{attribute–sentiment level}, directly complementing absent or insufficient fine-grained preference signals. 

\textsc{Mosaic} proceeds in three components organized as a \textbf{cascaded framework}. First, we extract attribute-sentiment pairs from individual user reviews and aggregate them across neighbor users to \textbf{formulate meta-reviews}. Then, a multi-gate MoE (MMoE) structure is designed to jointly perform \textbf{rating prediction} and \textbf{attribute-sentiment label prediction}, where the latter provides explicit fine-grained preference signals beyond serving as an auxiliary explainability task. Finally, a \textbf{personalized attention module} conditions on the target user to selectively highlight relevant attributes from the meta-review, yielding refined rating predictions and attribute-level explanations. 

In summary, our contributions are outlined as follows:
\begin{itemize}[leftmargin=12pt, itemsep=2pt, topsep=2pt, parsep=0pt]
    \item Motivated by the meta-review in the academic peer review process, we introduce a new perspective, \textbf{attribute-level meta-reviews} aggregated from neighbor users, to address UGC \textbf{sparsity and incompleteness}.
    \item We propose \textsc{Mosaic}, a novel multi-task review-based recommendation framework that incorporates a \textbf{personalized attention module}, selectively highlighting the attributes most relevant to the target user for refined rating prediction and attribute-level explanation.
    \item Extensive experiments on four public benchmark datasets demonstrate the superiority of \textsc{Mosaic} over state-of-the-art baselines. Further ablation studies verify the effectiveness of each component, and additional analyses confirm consistent gains for users with limited interaction history.
\end{itemize}

\section{Related Work}
\subsection{Explainable Recommender Systems}
Explainable recommender systems provide textual or attribute-level justifications. Existing methods either extract existing phrases or reviews \cite{wang2022graph,pugoy2021unsupervised}, or generate natural-language explanations with RNN-, Transformer-, PLM-, or LLM-based models \cite{Li_2017,li2021personalized,li2023personalized,ariza2023towards,li2023bprompt,zhou2024enhancing,Zhang_2024}. Recent hybrid methods further ground generation on retrieved reviews or attributes to reduce hallucination \cite{cheng2023explainable,10096389}. However, these methods still rely on available review content and thus struggle when reviews are missing or incomplete.

\subsection{Review-based Recommendation}
Review-based recommendation uses reviews to improve rating prediction beyond ID-based models \cite{5197422}. Prior work encodes review semantics with CNNs \cite{kim2016convolutional,catherine2017transnets}, attention mechanisms \cite{chen2018neural,seo2017interpretable,liu2019daml,wu2019context}, graph or contrastive learning \cite{wang2023learning,Liu_2025,shuai2022review,yang2023based,wang2023multi,wei2024multi}, and LLM-based aspect extraction \cite{li2023prompt}. Nevertheless, most methods depend on individually coarse review-level signals, leaving UGC sparsity and attribute-level incompleteness insufficiently addressed. We provide a more comprehensive discussion of related studies in Appendix~\ref{related_work}.

\section{Motivation}
\subsection{LLM-Based Attribute-Sentiment Extraction}
\label{sec:llm_extraction}
LLMs have demonstrated strong capabilities in semantic understanding and structured information extraction. Leveraging these capabilities, we employ an LLM to extract concise attribute-sentiment pairs from unstructured review texts, which serve as both interpretable building blocks for explanations and semantic features for review-based recommendation.

Let $s_{u,i}$ denote a review written by user $u$ on item $i$. Given a predefined domain-specific attribute set $\mathcal{A} = \{a_1, a_2, \dots, a_A\}$ with $A$ attributes, we 
define the extraction function performed by the LLM as:
\setlength{\abovedisplayskip}{4pt}
\setlength{\belowdisplayskip}{4pt}
\[
f_{\text{LLM}}(s_{u,i}, \mathcal{A}) \rightarrow \{(a_j, o_j) \mid a_j \in \mathcal{A}, o_j \in \mathcal{O} \cup \Phi \},
\]
where $o_j$ denotes the extracted sentiment for attribute $a_j$ and $\mathcal{O}$ is the vocabulary of possible sentimental phrases (limited to 3 words). If an attribute $a_j$ is not explicitly mentioned or lacks sentiment polarity in the review, the sentiment set is empty (represented by "N/A" in the natural language description).

To construct the attribute set $\mathcal{A}$ for each dataset, we reference domain-specific attribute taxonomies, guided by prior work such as FineRec~\cite{Zhang_2024}. In particular, the attribute list is constructed by selecting the most frequently mentioned features across all reviews, ensuring coverage and relevance. Then, the LLM is instructed using a carefully designed prompt to encourage structured and sentiment-aware extraction. In our paper, \textit{gpt-4.1-nano} is used for attribute-sentiment pairs extraction, the full prompt template is provided in Appendix~\ref{app:prompt}.






 Then, we adopt the sentiment analysis to leverage the extracted attribute-sentiment pairs. Specifically, we combine each attribute and its corresponding extracted sentiment as a new sentence: "\textit{The \{attribute\} is \{sentiment\}}", and use a fine-tuned pretrained BERT \cite{sanh2019distilbert} to do a sentiment classification with $C$ classes. In this paper, we set $C=4$,  which categorizes the sentiment into one of four classes: \emph{positive}, \emph{slightly positive}, \emph{slightly negative}, and \emph{negative}. 
Upon obtaining the sentiment scores, we can obtain the sentiment label $\mathbf{Y}_{u,i} \in {\{0,1\}}^{A \times C}$ for each user $u$ and item $i$.

Notably, viewed as a superior alternative to traditional extraction methods\cite{zhou2024enhancing}, the LLM does not participate in model training or inference and does not introduce additional computational cost(see Appendix~\ref{llm-analysis}). 

\subsection{Incompleteness of User Review}
\label{motivation}

In contrast to sparsity, where a review is entirely missing, incompleteness arises when a user, despite writing a review, does not mention all the attributes they considered. Although both phenomena degrade recommendation quality, 
\textbf{incompleteness of the review} has been largely overlooked in prior work~\cite{lei2025feature, vartak2017meta}. In this section, we empirically investigate this overlooked problem and motivate our solution.

\begin{figure}[!t]
    \centering
    \includegraphics[width=0.95\linewidth]{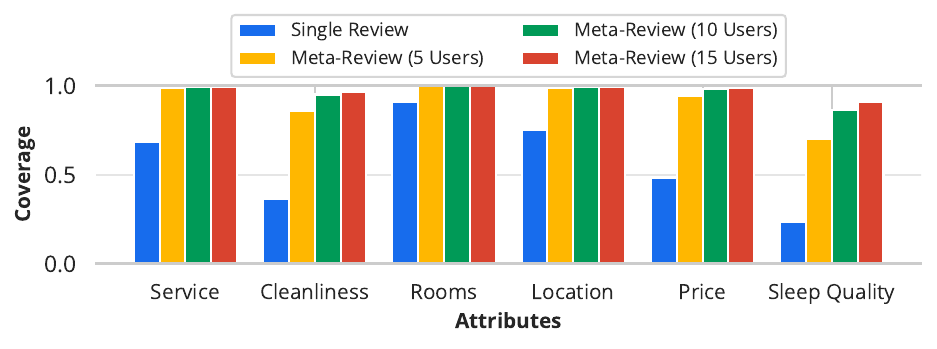}
    \caption{Coverage of attributes in single review and meta-reviews on the Tripadvisor dataset.}
    \label{fig:attr_incompl}
\end{figure}

To validate the existence of review incompleteness, we first extracted attribute-sentiment pairs from the \textit{TripAdvisor} dataset using six predefined attributes: \textit{Service, Cleanliness, Rooms, Location, Price, and Sleep Quality}. As shown in Figure \ref{fig:attr_incompl}, our analysis reveals that half of the attributes had a coverage rate below 50\% across all reviews, confirming that single user reviews are often incomplete. This phenomenon is also observable at the individual level. For example, Figure \ref{fig:user_review} illustrates a case where a user gave the same hotel an identical rating on two occasions but mentioned a different set of attributes each time. This suggests that a user's written feedback may not fully capture their comprehensive preferences.

In academic peer review, a single reviewer's opinion may not capture a paper's full quality, motivating the meta-review process that aggregates multiple reviewers~\cite{bhatia2020metagen}. Inspired by this, we aggregate reviews from neighbor users to 
form a more complete representation of preferences. Our exploratory analysis, depicted in Figure \ref{fig:attr_incompl}, demonstrates that aggregating reviews from as few as 10 neighbors can increase attribute coverage to over 80\%, effectively counteracting incompleteness. This aggregation strategy also addresses the sparsity problem by generating preference signals for interactions where the target user left no review.

\enlargethispage{1\baselineskip}
\section{Methodology}
\subsection{Model Overview}
\begin{figure}
    \centering
    \includegraphics[width=1.0\linewidth]{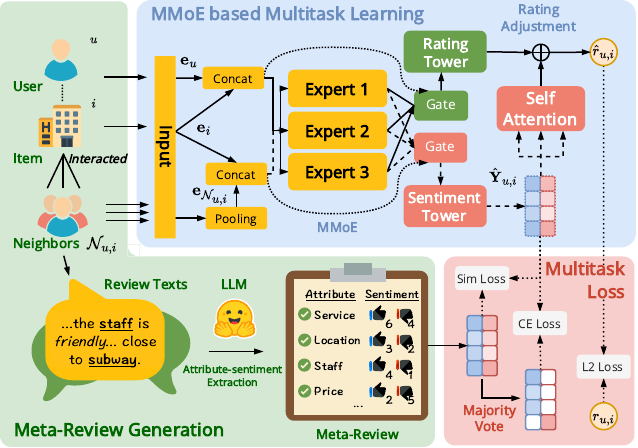}
    \caption{Overview of \textsc{Mosaic}. (1) \textbf{Meta-review Generation} LLM extracts attribute-sentiment pairs from neighbor users' 
reviews. (2) \textbf{MMoE-Based Multitask Learning}: MMoE jointly models rating and attribute-sentiment prediction, followed by a personalized attention module that conditions on the target user to refine ratings. (3) \textbf{Multitask Loss}: regression, cross-entropy, and similarity loss jointly supervise training.}
    \label{fig:model}
\end{figure}

\textsc{Mosaic} frames recommendation as a multi-task learning problem that jointly performs \textbf{rating prediction} and \textbf{attribute-sentiment label prediction} over both user-level and group-level (meta-review) signals. To balance 
these two tasks and prevent negative transfer~\cite{wei2023expgcn}, we adopt a \emph{Multi-gate Mixture-of-Experts} 
(MMoE)~\cite{ma2018modeling} architecture with task-specific towers. A \textbf{personalized attention module} then calibrates the predicted attribute-sentiments to each target user and refines the rating prediction. The full pipeline is illustrated in Figure~\ref{fig:model}.

\subsection{Problem Formulation}
In this paper, we formulate our framework as a multi-task learning problem that jointly performs \textbf{rating prediction} and \textbf{attribute-sentiment label prediction}. 

\begin{definition}[Rating Prediction]
Suppose the user set $\mathcal{U}$ and item set $\mathcal{I}$, given a user $u \in \mathcal{U}$ and item $i \in \mathcal{I}$, our task is to predict a rating $\hat{r}_{u, i}$ to align with the groundtruth rating $r_{u, i}$.
\end{definition}

\begin{definition}[Attribute-sentiment Label Prediction]
Given a user $u \in \mathcal{U}$ and item $i \in \mathcal{I}$, we predict a sentiment label $\hat{\mathbf{Y}}_{u,i} \in {\{0,1}\}^{A \times C}$, which consists of $A$ attributes and $C$ sentiment types. The prediction is going to be aligned with the groundtruth sentiment label $\mathbf{Y}_{u,i}$ extracted from user review $s_{u,i}$, only when $s_{u,i}$ is available. 
\end{definition}



\subsection{Construction of User Meta-Review}
\label{meta-review}
\noindent\textbf{Locating Neighbor Users}. For a target user $u$ and a target item $i$, we identify candidate neighbors as users who 
have reviewed the same item, optionally augmented with the target user's social friends. From this candidate set, we randomly sample $N=15$ users as neighbor users $\mathcal{N}_{u,i}$. We intentionally do \emph{not} filter neighbors by 
sentiment alignment with the target user since potential sentiment heterogeneity is mitigated by downstream majority voting. Candidate selection aims at broad attribute coverage, while \textbf{preference alignment is deferred to the personalized attention module}, which selectively 
emphasizes neighbor sentiments that match the target user's preferences. As illustrated in Figure \ref{fig:attr_incompl}, randomly sampling $N=15$ users is already sufficient to achieve adequate diversity at the attribute level. 



\noindent \textbf{Generating User Meta-Review.}  Given the sentiment label $\mathbf{Y}_{u,i} \in {[0,1]}^{A \times C}$ for user $u$ and item $i$, we can derive the meta-review attribute-sentiment vector $\mathbf{V}_{\mathcal{N}_{u,i}} \in \mathbb{R}^{A \times C}$ for the user neighbor set $\mathcal{N}_{u,i}$ through mean pooling, and obtain the meta-review attribute-sentiment labels $\mathbf{Y}_{\mathcal{N}_{u,i}} \in [0,1]^{A \times C}$ for the user neighbor set $\mathcal{N}_{u,i}$ through majority voting on each attribute respectively:

\vspace{-8pt}
\begin{equation}
\scalebox{0.9}{$\mathbf{V}_{\mathcal{N}_{u,i}} = \frac{\sum_{u \in \mathcal{N}_{u,i}} \mathbf{Y}_{u,i}}{|\mathcal{N}_{u,i}|}  , \: \: \mathbf{Y}_{\mathcal{N}_{u,i}} = \text{MajorVote}_{a\in \mathcal{A}}(\mathbf{V}_{\mathcal{N}_{u,i}})$},
\label{eq:neighbor}
\end{equation}
where $\text{MajorVote}_{a\in \mathcal{A}}$ denotes selecting the sentiment labels with the maximum value in each attribute $a$, thereby helps reduce noise and mitigate potential sentiment heterogeneity across users. The derived meta-review sentiment representations, $\mathbf{V}_{\mathcal{N}_{u,i}}$ and $\mathbf{Y}_{\mathcal{N}_{u,i}}$, are jointly used as supervision signals for the sentiment analysis of the neighbor users, which helps balance collective signals and reduces the influence of extreme or inconsistent user opinions.


\subsection{Model Architecture}

\noindent \textbf{\textit{Input Representation.}} 
The rating prediction task uses only the target user embedding $\mathbf{e}_u$ and item embedding $\mathbf{e}_i$, while the attribute-sentiment 
prediction task additionally incorporates an aggregated neighbor embedding:
\par\nobreak\vspace{-8pt}
\begin{equation}
   \scalebox{0.9}{$ \mathbf{e}_{\mathcal{N}_{u,i}} = \frac{1}{|\mathcal{N}_{u,i}|} \sum_{g_i \in \mathcal{N}_{u,i}} \mathbf{e}_{g_i}$,}
\end{equation}
where $\mathcal{N}_{u,i}$ is the located neighbor users. Thus, For task $t \in \{r,o\}$, where $r$ denotes rating prediction and $o$ denotes meta-review sentiment prediction, the input vectors are:
\par\nobreak\vspace{-8pt}
\begin{equation}
\scalebox{0.9}{$\mathbf{e}^{r} = \mathbf{e}_u || \mathbf{e}_i \in \mathbb{R}^{2d}, 
\mathbf{e}^{o} = \mathbf{e}_{\mathcal{N}_{u,i}} || \mathbf{e}_i \in \mathbb{R}^{2d}$},
\end{equation}
where $||$ denotes the concatenation operation.



\subsubsection{Multi-Task Integration of Meta-Review}
To balance the two tasks while mitigating negative transfer between individual reviews and meta-review signals, we adopt the MMoE framework with shared experts and task-specific gating networks. Each gate dynamically weights the experts' contributions for its task, creating functionally distinct feature spaces for the disparate input representations of the two tasks. The components of the MMoE in \textsc{Mosaic} are detailed as follows:

\noindent \textbf{\textit{Expert Networks.}} 
For each expert $k = 1,\dots,K$, we employ a Transformer encoder \cite{li2021personalized} to capture contextualized representations. The expert transformation is given by:
\par\nobreak\vspace{-8pt}
\begin{equation}
  \scalebox{0.9}{$  \mathbf{x}_{k}^t = \phi(\mathbf{W}^{\mathit{I}}_k\mathbf{h}^{t} + \mathbf{b}^{\mathit{I}}_k),\:\:
    \mathbf{z}_{k}^t = \mathrm{Transformer}_k(\mathbf{x}_{k}^t)$},
\end{equation}
where $\mathbf{h}^t$ is the input vector after the feature representation layer, $\mathbf{W}^{\mathit{I}}_k, \mathbf{b}^{\mathit{I}}_k$ are input projection parameters for expert $k$, $\mathbf{x}_{k}^t$ is the projected input to expert $k$, and $\mathbf{z}_{k}^t$ is its contextualized output.

\noindent \textbf{\textit{Task-specific Gating.}} 
The MMoE module adaptively assigns expert contributions to each task through a softmax gating mechanism. For task $t \in \{r,o\}$, the output is:
\par\nobreak\vspace{-8pt}
\begin{equation}
\scalebox{0.9}{$ \mathbf{m}^{t} = \sum_{k=1}^K \alpha_k^t \mathbf{z}^t_k$},
\end{equation}
where the gating weights $\alpha_k$ for each expert $k$ are defined as: 
\par\nobreak\vspace{-8pt}
\begin{equation}
  \scalebox{0.9}{$ \alpha_k = \frac{\exp\left((\mathbf{w}_k^{t})^T \mathbf{h}^t + b_k^{t}\right)}{\sum_{j=1}^{K} \exp\left((\mathbf{w}_j^{t})^T \mathbf{h}^t + b_j^{t}\right)}$},
\end{equation}
where $\mathbf{w}_k^{t}, b_k^{t}$ are gating parameters for expert $k$ and task $t$.

\noindent \textbf{\textit{Task-specific Towers.}} Given the aggregated representations $\mathbf{m}^r$ and $\mathbf{m}^o$ from the MMoE 
module, two task-specific towers produce the final outputs of each task. The rating tower applies an MLP on $\mathbf{m}^r$ to 
predict the initial rating $\hat{r}^r$, while the sentiment tower applies an MLP followed by a softmax on $\mathbf{m}^o$ to produce attribute-wise sentiment distributions $\hat{\mathbf{y}}_a \in \mathbb{R}^C$ for each attribute $a = 1, \ldots, A$.




\subsubsection{Personalization of Meta-Review Signals}  
While meta-reviews aggregate sentiment signals across multiple neighbor users, the target user's specific preferences may emphasize only a subset of these attributes. We therefore introduce a user-specific personalization step that selectively highlights the attributes most relevant to the target user, ensuring that the aggregated neighbor signals are calibrated 
to individual preferences rather than treated uniformly.

Specifically, the personalization step is implemented as a self-attention module that takes the rating features $\mathbf{m}^r$ as the query and predicted attribute-sentiment signals as keys and values, producing user-conditioned weights that emphasize attributes aligned with the target user's preferences and suppress less relevant ones. We then collect all attribute predictions into $\hat{\mathbf{Y}} = [\hat{\mathbf{y}}_{a_1}, \dots, 
\hat{\mathbf{y}}_{a_A}]$ and project them through a sentiment embedding space via a learnable matrix $\mathbf{W}^S \in \mathbb{R}^{C \times d}$, followed by a self-attention layer 
to obtain the pooled representation:
\par\nobreak\vspace{-8pt}
\begin{equation}
   \scalebox{0.9}{$ \mathbf{p} = \text{Self-attention}(\hat{\mathbf{Y}} \mathbf{W}^{\mathit{S}}, \mathbf{m}^{r}).$}
    \label{eq:attn}
\end{equation}

The adjustment rating score $\hat{r}^o$ is obtained by combining the sentiment representation $\mathbf{p}$ with the rating features $\mathbf{m}^{r}$ through non-linear perceptron layer $\Gamma(\cdot)$ as the fusion gate:
\par\nobreak\vspace{-8pt}
\begin{equation}
    \scalebox{0.9}{$\hat{r}^o = MLP\left(\mathbf{m}^{r} \cdot \Gamma(\mathbf{m}^{r}|| \mathbf{p}) + \mathbf{p} \cdot (1-\Gamma(\mathbf{m}^{r}|| \mathbf{p}))\right).$}
\end{equation}
The attention module adaptively weighs user-item representations and meta-review sentiment, enabling the model to refine the final rating prediction with contextual signals.

\subsection{Loss Functions}

The final rating combines the initial prediction $\hat{r}^r_{u,i}$ and the meta-review guided adjustment $\hat{r}^o_{u,i}$, 
supervised by squared error against the ground truth $r_{u,i}$:
\par\nobreak\vspace{-8pt}
\begin{equation}
\scalebox{0.9}{$\mathcal{L}_{rating} = \frac{1}{|\mathcal{T}|} \sum_{(u,i) \in \mathcal{T}} \big(\hat{r}_{u,i} - r_{u,i}\big)^2$},
\end{equation}
where $\mathcal{T}$ is the training set.


In the meta-review attribute-sentiment label prediction task, the user-level sentiment label $\mathbf{Y}_{u,i}$ supervises $\hat{\mathbf{Y}}_{\mathcal{N}_{u,i}}$ at the user-level to maintain label consistency, while the ground-truth group-level labels $\mathbf{Y}_{\mathcal{N}_{u,i}}$ from meta-reviews via majority vote (Section \ref{meta-review}) provide broader supervision. As $\mathbf{Y}_{\mathcal{N}_{u,i}}$ is derived from the meta-review sentiment vector $\mathbf{V}_{\mathcal{N}_{u,i}}$, the original sentiment distribution preserves richer and more nuanced information. Therefore, we design a joint loss that includes both the classification loss of the sentiment label $\mathbf{Y}_{\mathcal{N}_{u,i}}$ and the similarity of the meta-review sentiment vector $\mathbf{V}_{\mathcal{N}_{u,i}}$, controlled by parameters $\eta_{cls}$ and $\eta_{sim}$. This joint supervision strategy enables the model to learn from consensus signals while preserving sentiment variability among neighbors, reducing the risk of overfitting to hard labels and limiting the impact of outlier users. The joint loss for meta-review sentiment prediction $\mathcal{L}_{meta}$ is given by:


\par\nobreak\vspace{-8pt}
\begin{equation}
  \scalebox{0.75}{$\displaystyle
  \begin{aligned}
 \mathcal{L}_{meta} = -\frac{1}{|\mathcal{T}|} \sum_{(u,i) \in \mathcal{T}} \eta_{cls}& \big(\mathbf{Y}_{u,i} \log \hat{\mathbf{Y}}_{\mathcal{N}_{u,i}} + \mathbf{Y}_{\mathcal{N}_{u,i}} \log \hat{\mathbf{Y}}_{\mathcal{N}_{u,i}}) \\ +\ \eta_{sim}&\   \text{Sim}\big(\hat{\mathbf{Y}}_{\mathcal{N}_{u,i}} \mathbf{W}^{\mathit{S}}, \mathbf{V}_{\mathcal{N}_{u,i}} \mathbf{W}^{\mathit{S}}).
  \end{aligned}
  $}
  \label{lossmeta}
\end{equation}
Finally, we can obtain the final loss function combination: $ \mathcal{L} = \mathcal{L}_{rating} + \mathcal{L}_{meta}.$
For the similarity function $\text{Sim}(\cdot)$, we adopt the cosine similarity.
The settings of hyperparameters $\eta_{cls}$ and $\eta_{sim}$ will be discussed in Section~\ref{sec:ablation}.

\section{Experiments}
\begin{table*}[t]
  \centering
  \scriptsize
  \caption{Statistics of the datasets.}
  \renewcommand{\arraystretch}{0.85}
  \setlength{\tabcolsep}{6pt}
  \label{dataset}
\begin{tabular}{@{}lcccc@{}}
\toprule
\textbf{Dataset} & \textbf{\# users} & \textbf{\# items} & \textbf{\# reviews} & \textbf{Attributes list} \\ \midrule
\textbf{Yelp} & 19,348 & 9,020 & 103,022 & \begin{tabular}[c]{@{}c@{}}Ambience, Cleanliness, Food, Location, Parking, Price, Service\end{tabular} \\
\textbf{TripAdvisor} & 3,255 & 6,261 & 104,570 & \begin{tabular}[c]{@{}c@{}}Service, Cleanliness, Rooms, Location, Price, Sleep Quality\end{tabular} \\
\textbf{Amazon Beauty} & 12,784 & 5,190 & 32,506 & \begin{tabular}[c]{@{}c@{}}Color, Effectiveness, Price, Ingredients, Scent, Size\end{tabular} \\
\textbf{Amazon Sports} & 32,374 & 24,122 & 140,603 & \begin{tabular}[c]{@{}c@{}}Brand, Comfort, Functionality, Material, Price, Quality, Size\end{tabular} \\ \bottomrule
\end{tabular}
\end{table*}


\subsection{Datasets and Baselines}
\noindent\textbf{Datasets.} 
We conduct experiments on four publicly available datasets spanning diverse application domains, which are \textit{Yelp} (restaurants), \textit{TripAdvisor} (hotels), \textit{Amazon Beauty}, and \textit{Amazon Sports}. Each record contains a user ID, an item ID, a numeric rating, and a corresponding review text (can be empty), annotated with attribute-sentiment pairs as described in Section~\ref{sec:llm_extraction}. We filter users with fewer than three rated items. Data statistics and predefined attributes are presented in Table~\ref{dataset}. Full descriptions and preprocessing details are in Appendix~\ref{app:datasets}.

\noindent\textbf{Baselines.}
We compare \textsc{Mosaic} against several state-of-the-art baselines, covering classical matrix factorization (\textbf{PMF}~\cite{NIPS2007_d7322ed7} and \textbf{SVD++}~\cite{10.1145/1401890.1401944}), review-based recommendation models (\textbf{RGCL}~\cite{shuai2022review}, \textbf{APH}~\cite{Liu_2025}), explainable recommendation models (\textbf{NETE}~\cite{li2020towards}, \textbf{PETER}~\cite{li2021personalized}, \textbf{PEPLER}~\cite{li2023personalized}, \textbf{CER}~\cite{raczynski2023problem}, \textbf{SERMON}~\cite{liao2025aspect}), and a LLM-based recommender (\textbf{LLMRec}~\cite{wei2024wsdm}). For the LLM-based baseline, we replace its original LLM backbones with \textit{gpt-4.1-nano} to ensure fair comparison with \textsc{Mosaic}'s extraction pipeline. Detailed descriptions are provided in 
Appendix~\ref{app:baselines}.

\begin{table*}[t]
\scriptsize
\centering
\renewcommand{\arraystretch}{0.7}
\setul{1pt}{0.4pt}
\setlength{\tabcolsep}{3pt}
\caption{Performance on rating prediction. In each row, the best 
result is in \textbf{bold} and the best baseline is \ul{underlined}. 
$^*$ indicates statistical significance with $p<0.01$. }
\label{tab:rmse_mae}
\begin{tabular}{ll cc cccccccc c}
\toprule
\textbf{Dataset} & \textbf{Metric} 
& PMF & SVD++ 
& RGCL & NETE & PETER & PEPLER & CER & SERMON & APH & LLMRec 
& \textbf{\textsc{Mosaic}} \\
\midrule
\multirow{2}{*}{\textbf{\ \ \ Yelp}} 
& RMSE & 0.9866 & 0.9405 & 0.9086 & 1.0529 & 0.9454 
  & \ul{0.8592} & 0.8652 & 1.0871 & 0.8504 & 1.1930 
  & \textbf{0.8239}$^*$ \\
& MAE  & 0.7119 & 0.7284 & 0.6951 & 0.7650 & 0.7064 
  & \ul{0.5854} & 0.6634 & 0.8142 & 0.6576 & 0.8975 
  & \textbf{0.5708} \\
\cmidrule(lr){1-13}
\multirow{2}{*}{\makecell[c]{\textbf{Trip-} \\ \textbf{Advisor}}}
& RMSE & 0.8777 & 0.9010 & \ul{0.7798} & 0.7852 & 0.8058 
  & 0.7799 & 0.8212 & 0.8936 & 0.7826 & 1.0238 
  & \textbf{0.6759}$^*$ \\
& MAE  & 0.6769 & 0.7060 & 0.5989 & \ul{0.5919} & 0.6329 
  & 0.6160 & 0.6406 & 0.6913 & 0.6055 & 0.8231 
  & \textbf{0.5303}$^*$ \\
\cmidrule(lr){1-13}
\multirow{2}{*}{\makecell[c]{\textbf{Amazon} \\ \textbf{Beauty}}}
& RMSE & 1.1073 & 0.8886 & \ul{0.8565} & 0.9884 & 1.1231 
  & 0.9613 & 1.1236 & 1.2940 & 0.8574 & 1.3176 
  & \textbf{0.7840}$^*$ \\
& MAE  & 0.8618 & 0.7124 & 0.6717 & 0.6774 & 0.8884 
  & 0.6650 & 0.8616 & 0.9141 & \ul{0.6452} & 0.7512 
  & \textbf{0.5400}$^*$ \\
\cmidrule(lr){1-13}
\multirow{2}{*}{\makecell[c]{\textbf{Amazon} \\ \textbf{Sports}}}
& RMSE & 0.9659 & 1.4380 & \ul{0.8866} & 0.9181 & 0.9360 
  & 0.9112 & 0.9375 & 1.0524 & 0.9383 & 1.4580 
  & \textbf{0.8224}$^*$ \\
& MAE  & 0.6912 & 1.0829 & \ul{0.5901} & 0.6478 & 0.6745 
  & 0.6493 & 0.6823 & 0.7473 & 0.6840 & 1.0326 
  & \textbf{0.4909}$^*$ \\
\bottomrule
\end{tabular}
\end{table*}
\begin{table*}[t]
\scriptsize
\centering
\renewcommand{\arraystretch}{0.7}
\setul{1pt}{0.4pt}
\setlength{\tabcolsep}{2pt}
\caption{Performance on meta-review sentiment prediction. In each 
column, the best result is in \textbf{bold} and the best baseline 
is \ul{underlined}. $^*$ indicates statistical significance with 
$p<0.01$. Acc., F1, Prec., Rec. denote Accuracy, F1 Score, Precision, 
and Recall respectively.}
\label{tab:opinion_prediction}
\begin{tabular}{l cccc cccc cccc cccc}
\toprule
& \multicolumn{4}{c}{\textbf{Yelp}} 
& \multicolumn{4}{c}{\textbf{TripAdvisor}} 
& \multicolumn{4}{c}{\textbf{Amazon Beauty}} 
& \multicolumn{4}{c}{\textbf{Amazon Sports}} \\
\cmidrule(lr){2-5} \cmidrule(lr){6-9} \cmidrule(lr){10-13} \cmidrule(lr){14-17}
\textbf{Method} 
& Acc. & F1 & Prec. & Rec. 
& Acc. & F1 & Prec. & Rec. 
& Acc. & F1 & Prec. & Rec. 
& Acc. & F1 & Prec. & Rec. \\
\midrule
NETE   
& 0.5909 & 0.1912 & 0.1625 & 0.2321 
& 0.2773 & 0.1199 & 0.2487 & 0.2476 
& 0.5959 & 0.1874 & 0.2257 & 0.2500 
& 0.3913 & 0.1522 & 0.1320 & 0.2371 \\
PETER  
& 0.5977 & 0.2024 & 0.1701 & 0.2500 
& 0.3444 & 0.1835 & \ul{0.3685} & 0.2867 
& 0.6122 & 0.2232 & 0.3617 & 0.2668 
& 0.4225 & 0.1586 & 0.3956 & 0.2544 \\
PEPLER 
& 0.5970 & \ul{0.2535} & 0.2665 & \ul{0.2593} 
& \ul{0.3714} & \ul{0.2193} & 0.2799 & \ul{0.2973} 
& 0.5965 & 0.1868 & 0.1491 & 0.2500 
& 0.4203 & 0.1481 & 0.3141 & 0.2501 \\
CER    
& 0.6145 & 0.2475 & 0.2668 & 0.2580 
& 0.3632 & 0.1979 & 0.3702 & 0.2947 
& 0.5974 & 0.1896 & \textbf{0.4042} & 0.2513 
& 0.4226 & 0.1594 & \ul{0.4100} & 0.2549 \\
SERMON 
& \ul{0.6797} & 0.2030 & \ul{0.2908} & 0.2501 
& 0.3696 & 0.2066 & 0.3657 & 0.2969 
& \ul{0.6465} & \ul{0.2684} & \ul{0.3273} & \ul{0.2954} 
& \ul{0.4243} & \ul{0.1741} & 0.3534 & \ul{0.2613} \\
\midrule
\textbf{\textsc{Mosaic}} 
& \textbf{0.7064}$^*$ & \textbf{0.3488}$^*$ & \textbf{0.5217}$^*$ & \textbf{0.3335}$^*$ 
& \textbf{0.5463}$^*$ & \textbf{0.2877}$^*$ & \textbf{0.3769} & \textbf{0.3102} 
& \textbf{0.6529} & \textbf{0.3101}$^*$ & 0.3119 & \textbf{0.3277}$^*$ 
& \textbf{0.7720}$^*$ & \textbf{0.5160}$^*$ & \textbf{0.5446}$^*$ & \textbf{0.5201}$^*$ \\
\bottomrule
\end{tabular}
\end{table*}

\begin{table*}[htbp]
\centering
\scriptsize
\renewcommand{\arraystretch}{0.7}
\setul{1pt}{0.4pt}
\setlength{\tabcolsep}{3pt}
\caption{Rating prediction on Yelp and Amazon Sports across user groups partitioned by interaction count. \textbf{Improv.} denotes \textsc{Mosaic}'s relative gain over the best baseline. 
\textbf{Best} and \underline{runner-up} results are highlighted.}
\label{tab:cold_start}
\setlength{\tabcolsep}{4pt}
\begin{tabular}{llcccccccc}
\toprule
\textbf{Dataset} & \textbf{Group (\# Samples)} & \textbf{Metric} 
& \textbf{\textsc{Mosaic}} & RGCL & PETER & PEPLER & APH & LLMRec 
& \textbf{Improv.} \\
\midrule
\multirow{6}{*}{\textbf{Yelp}} 
& \multirow{2}{*}{Sparse ($\leq 1$) (13{,}604)} 
  & RMSE & \textbf{0.8904} & 1.1621 & 1.1268 & 1.3054 
  & \underline{1.0581} & 1.4567 & \textbf{15.85\%} \\
& & MAE  & \textbf{0.5902} & 0.8682 & 0.8793 & 1.0589 
  & \underline{0.8477} & 1.1338 & \textbf{30.38\%} \\
\cmidrule(lr){2-10}
& \multirow{2}{*}{Moderate (2--5) (15{,}070)} 
  & RMSE & \textbf{0.7924} & 1.0252 & 0.9872 & 1.2083 
  & \underline{0.9497} & 1.2383 & \textbf{16.56\%} \\
& & MAE  & \textbf{0.5902} & \underline{0.7426} & 0.7624 & 0.9762 
  & 0.7482 & 0.9179 & \textbf{20.52\%} \\
\cmidrule(lr){2-10}
& \multirow{2}{*}{Active ($>5$) (8{,}749)} 
  & RMSE & \textbf{0.7860} & 0.9244 & 0.8854 & 1.1194 
  & \underline{0.8478} & 0.9819 & \textbf{7.29\%} \\
& & MAE  & \textbf{0.5984} & \underline{0.6639} & 0.6865 & 0.9017 
  & 0.6648 & 0.7561 & \textbf{9.87\%} \\
\midrule
\multirow{6}{*}{\textbf{Sports}} 
& \multirow{2}{*}{Sparse ($\leq 1$) (14{,}701)} 
  & RMSE & \textbf{0.7050} & 0.9436 & 0.9890 & 0.9635 
  & \underline{0.9119} & 1.6336 & \textbf{22.69\%} \\
& & MAE  & \textbf{0.4425} & \underline{0.6053} & 0.6156 & 0.7132 
  & 0.6740 & 1.2683 & \textbf{26.90\%} \\
\cmidrule(lr){2-10}
& \multirow{2}{*}{Moderate (2--5) (11{,}162)} 
  & RMSE & \textbf{0.6075} & 0.8980 & 0.9601 & 0.8993 
  & \underline{0.8743} & 1.4076 & \textbf{30.52\%} \\
& & MAE  & \textbf{0.3905} & \underline{0.5496} & 0.6022 & 0.6514 
  & 0.6291 & 0.9617 & \textbf{28.95\%} \\
\cmidrule(lr){2-10}
& \multirow{2}{*}{Active ($>5$) (4{,}791)} 
  & RMSE & \textbf{0.5266} & \underline{0.7693} & 0.8241 & 0.7831 
  & 0.7726 & 1.2621 & \textbf{31.55\%} \\
& & MAE  & \textbf{0.3577} & \underline{0.4666} & 0.5373 & 0.5771 
  & 0.5469 & 0.8281 & \textbf{23.34\%} \\
\bottomrule
\end{tabular}
\end{table*}
\begin{figure}[!t]
\centering

\begin{minipage}[c]{\linewidth}
\centering
\includegraphics[width=0.9\linewidth]{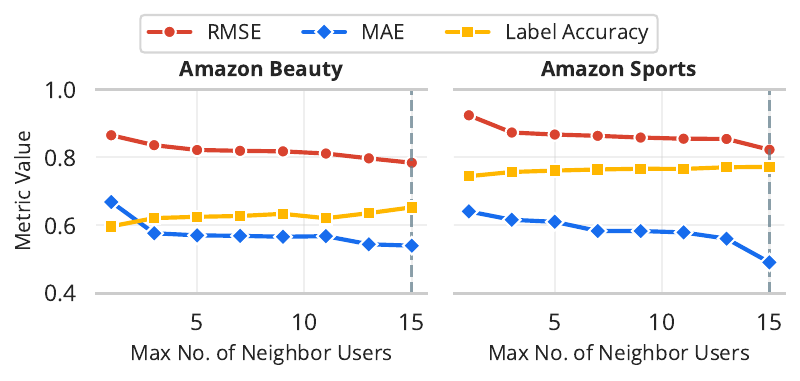}
\captionof{figure}{Performance with different numbers of neighbor users.}
\label{fig:groups}
\end{minipage}

\vspace{6pt}

\begin{minipage}[c]{\linewidth}
\centering
\includegraphics[width=0.9\linewidth]{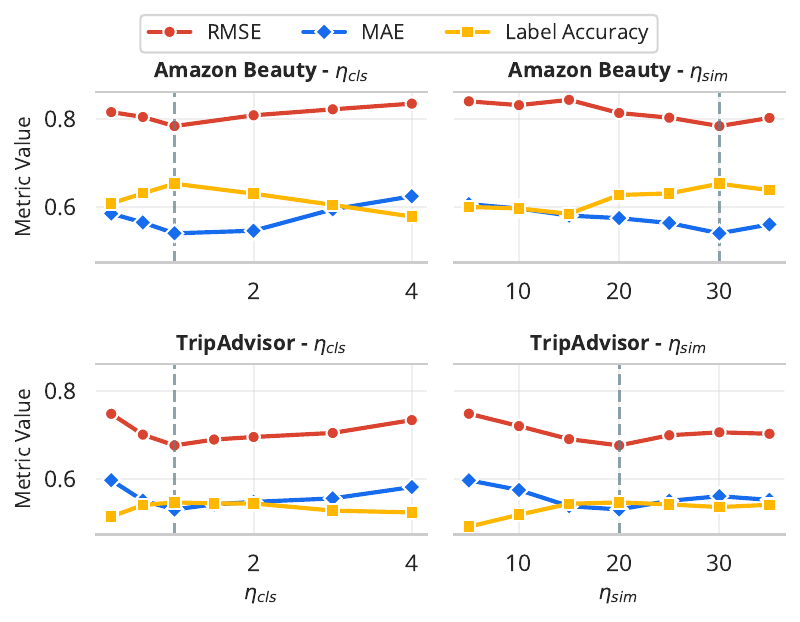}
\captionof{figure}{Performance with different $\eta_{cls}$ and $\eta_{sim}$.}
\label{fig:parameter_eta}
\end{minipage}

\end{figure}

\begin{table}
\centering
\scriptsize
\renewcommand{\arraystretch}{0.7}
\setul{1pt}{0.4pt}
\setlength{\tabcolsep}{2pt}
\caption{Ablation study.}
  \label{tab:loss}
  \begin{tabular}{@{}l|ccc|ccc@{}}
  \toprule
  \multicolumn{1}{c|}{\multirow{2}{*}{\textbf{Model}}} & \multicolumn{3}{c|}{\textbf{Beauty}} & \multicolumn{3}{c}{\textbf{TripAdvisor}} \\ \cmidrule(l){2-7} 
  \multicolumn{1}{c|}{} & RMSE & MAE & Accuracy & RMSE & MAE & Accuracy \\ \midrule
 w/o Neighbor Users & 0.9331 & 0.6778 & 0.5887 & 0.7974 & 0.6167 & 0.5168 \\
 w/o Attention & 0.9252 & 0.6452 & 0.6372 & 0.7857 & 0.6109 & 0.5003 \\
  $\eta_{cls},\eta_{sim}=0$ & 0.9721 & 0.5663 & 0.3073 & 0.7249 & 0.5734 & 0.4188 \\
  $\eta_{sim}=0$ & 0.8533 & 0.6007 & 0.5283 & 0.7172 & 0.5681 & 0.4969 \\
  $\eta_{cls}=0$ & 0.8485  &0.6051  & 0.6289 & 0.7154 & 0.5706 & 0.5361 \\ \midrule
  Full Model & \textbf{0.7840} & \textbf{0.5400} & \textbf{0.6529} & \textbf{0.6759} & \textbf{0.5303} & \textbf{0.5463} \\ \bottomrule
  \end{tabular}
\end{table}

\subsection{Overall Performance}
We evaluate rating prediction with RMSE and MAE, and attribute-sentiment prediction with Accuracy, F1 Score, Precision, and Recall. Full experimental settings, including 
data split, training and evaluation details are in Appendix ~\ref{experimental_settings}.

\textbf{Recommendation Performance.} Table~\ref{tab:rmse_mae} reports the rating prediction performance on the four datasets. The results demonstrate that \textsc{Mosaic} consistently outperforms all comparative methods under both metrics across all datasets. In particular, our method achieves superior performance over classical collaborative filtering models such as PMF and SVD++, review-based methods such as RGCL and APH, joint rating--explanation models (NETE, PETER, PEPLER, CER, SERMON), a LLM-based recommender(LLMRec). Notably, \textsc{Mosaic} surpasses RGCL by aggregating neighborhood information at the \textbf{attribute-level} rather than the review level, directly addressing review incompleteness. Additionally, LLMRec's weaker performance indicates that LLM-driven graph augmentation alone is less effective for fine-grained rating regression. \textsc{Mosaic} focuses on user-side evidence and achieves consistent advantages on all four datasets.



\textbf{Meta-review Attribute-sentiment Prediction Performance.} For this task, we include only explanation-capable methods, namely NETE, PETER, PEPLER, CER, and SERMON. As shown in Table \ref{tab:opinion_prediction}, \textsc{Mosaic} achieves the best overall performance. While some baselines achieve slightly higher precision, their lower F1 and recall reveal a tendency to 
predict positive labels only when highly certain. \textsc{Mosaic} instead identifies a broader range of correct labels, yielding more balanced performance across metrics.


Beyond these performance gains, Appendix~\ref{llm-analysis} shows that \textsc{Mosaic} maintains competitive training speed and low memory consumption compared to baselines.

\subsection{Ablation and Sensitivity Analysis}
\label{sec:ablation}



We conduct ablation and sensitivity analysis to evaluate the contribution of each proposed component and the model's robustness to key hyperparameters. Table~\ref{tab:loss} shows that the full model consistently outperforms all variants, validating each proposed strategy. Specifically, we evaluate the performance of the following model variants:

\noindent\textbf{Effect of Neighbor Users.} 
Removing neighbor users causes the largest drop on both datasets (e.g., RMSE 0.6759 $\to$ 0.7974 on TripAdvisor). As shown in Figure~\ref{fig:attr_incompl}, without meta-reviews aggregated from neighbors, the model loses access to the broader attribute coverage and the cross-user evidence required to compensate for review incompleteness, which directly translates to weaker rating prediction and sentiment label accuracy.

\textbf{Effect of Number of Neighbor Users.}  
To examine the model's \emph{sensitivity} to the number of neighbors, Figure~\ref{fig:groups} illustrates that increasing the maximum number of neighbor users generally improves RMSE, MAE, and label accuracy. These results confirm that richer attribute coverage enhances both prediction tasks. This aligns with our Figure~\ref{fig:attr_incompl}, where 10--15 neighbors are sufficient to achieve over 80\% attribute coverage.

\noindent\textbf{Effect of Attention.}
Replacing the self-attention module with a uniform MLP deteriorates performance across all metrics, such as RMSE 0.7840 $\to$ 0.9252 on Beauty. The MLP treats all attributes 
in the meta-review uniformly, whereas the attention module learns user-specific weights that highlight relevant attributes and suppress less relevant ones. Without this personalization step, the group-level meta-review signals cannot be calibrated to the target user, leading to 
biased rating adjustment.

\noindent\textbf{Effect of Meta-Review Supervision.}
The two coefficients play complementary roles: $\eta_{cls}$ aligns predictions with majority-voted hard labels, while $\eta_{sim}$ preserves the fine-grained sentiment distribution that majority voting discards. Disabling either or both signals causes sentiment accuracy to collapse. Interestingly, MAE slightly improves when removing both compared to removing only one, reflecting the inherent competing tension between the two tasks. \emph{Beyond ablation,} we vary $\eta_{cls} \in [1,5]$ and $\eta_{sim} \in [5,35]$ (with a step size of 5), shown in Figure~\ref{fig:parameter_eta}. The results indicate that $\eta_{sim}$ requires a substantially larger weight than $\eta_{cls}$ to balance the soft-distribution alignment against the hard-label classification(e.g. $\eta_{cls}=1$, $\eta_{sim}=30$ on the \textit{Beauty} dataset).




\subsection{Performance on Sparse Users}
We further evaluate \textsc{Mosaic} on users with limited interaction history. With few review evidence, it leaves a challenging setting for preference modeling due to highly limited attribute coverage. 

\textbf{Setting.} We partition the test interactions of the Yelp and Amazon Sports datasets into three groups based on the number of training interactions associated per test user: \emph{Sparse} ($\leq 1$ training interaction), \emph{Moderate} ($2$--$5$), and \emph{Active} ($>5$). The \emph{Sparse} group corresponds to users with the minimum training interactions under our split. We compare \textsc{Mosaic} against the strongest baselines from Table~\ref{tab:rmse_mae} (RGCL, APH, PEPLER, PETER) and LLMRec, whose explicit focus on data sparsity makes it particularly relevant here.


\textbf{Results.} Table~\ref{tab:cold_start} shows \textsc{Mosaic} consistently outperforms all 
baselines across the three groups on both datasets. The advantage is particularly pronounced in the \emph{Sparse} group, where \textsc{Mosaic} reduces MAE by $30.38\%$ on Yelp and $26.90\%$ on Amazon Sports relative to the strongest baseline. This pattern strongly illustrates that meta-reviews aggregated from neighbor users provide the missing fine-grained signals when the target user's own evidence is minimal. The advantage persists in the \emph{Moderate} and \emph{Active} groups, suggesting that group-level evidence remains informative even when individual review history grows richer. These results validate our design motivation that aggregating attribute-level meta-reviews offers a consistent benefit across a 
wide spectrum of user activity levels. Notably, LLMRec shows weaker performance throughout, suggesting that LLM-driven graph augmentation alone may be insufficient for the attribute-level reasoning required in review-based rating prediction.

\begin{figure}
    \centering
    \includegraphics[width=0.9\linewidth]{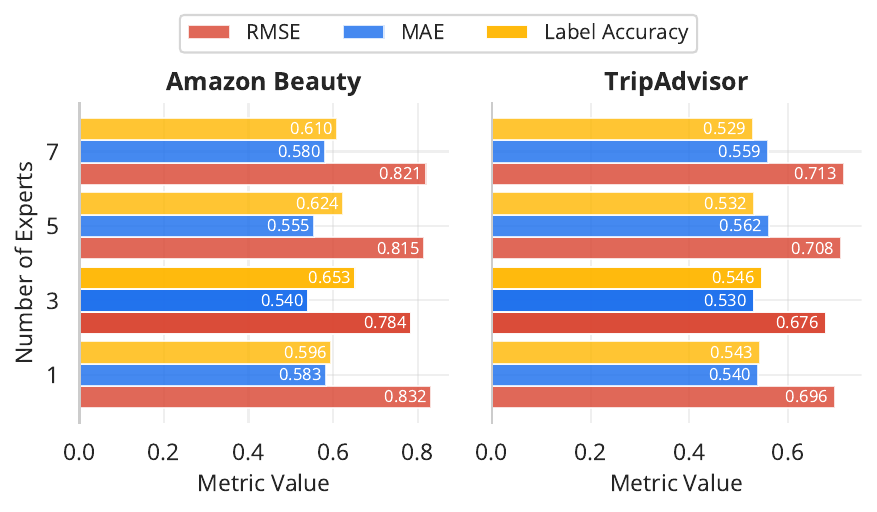}
    \caption{Performance with different number of experts $K$.}
    \vspace{-1em}
    \label{fig:experts}
\end{figure}

\begin{figure}
\centering
\includegraphics[width=0.9\linewidth]{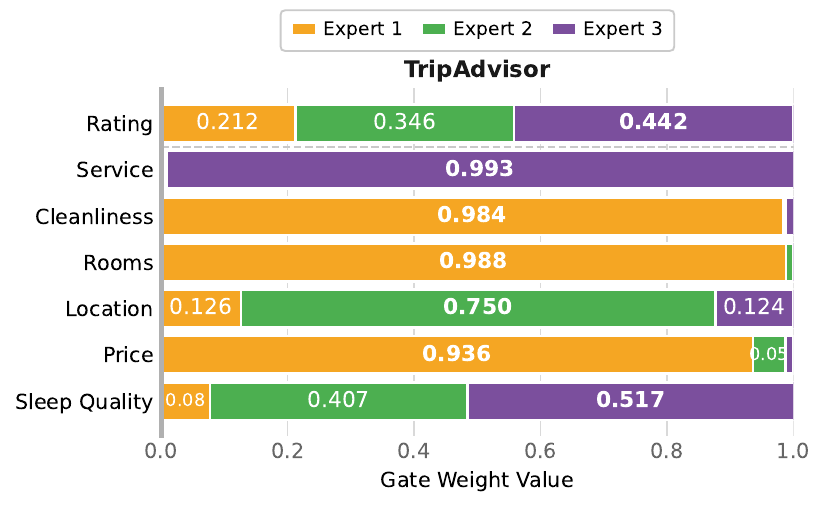}
\caption{Mean gate weights across experts at $K{=}3$ on TripAdvisor(13{,}027 test samples).}
\label{fig:gate_k3}
\end{figure}

\subsection{Number of Experts and Gating Behavior in MMoE}  
We analyze the impact of varying the number of experts in the MMoE module to account for the heterogeneity of task-specific inputs. As shown in Figure~\ref{fig:experts}, performance consistently improves as the number of experts $K$ increases from 1 to 3, while further increasing the number of experts leads to diminishing performance gains. The best trade-off for both rating prediction and attribute-sentiment prediction is observed when the number of experts is 3.

To understand this trend, we examine the learned gate distributions on TripAdvisor. As shown in Figure~\ref{fig:gate_k3}, at $K{=}3$ each attribute gate concentrates its weight on a dominant expert (e.g., Cleanliness 0.984 and Rooms 0.988 on Expert 1, Location 0.750 on Expert 2, Service 0.993 on Expert 3), while the rating gate distributes far more evenly (0.212 / 0.346 / 0.442). The two tasks thus learn markedly different routing patterns over the same shared experts, confirming that the task-specific gates perform differentiated, semantically meaningful routing rather than collapsing to a shared representation, which is precisely what a single shared bottom cannot produce by construction. Importantly, gating does not break down at larger $K$ and every attribute is still routed to a dominant expert. However, the extra experts do not form new specialists but instead fragment attributes that a single expert handled at $K=3$, so each expert receives a weaker, more diffuse training signal as the parameter count grows. This explains the slight degradation beyond three experts, with full $K=5/7$ distributions and analysis are provided in Table~\ref{tab:gate_k57} in the Appendix ~\ref{gate_distri}. Overall, these results confirm that $K{=}3$ offers the best balance between expert specialization and model capacity.


\vspace{6pt}

\vspace{-2pt}
\subsection{Case Study}
\label{case_study}
\begin{figure}
    \centering
    \includegraphics[width=0.9\linewidth]{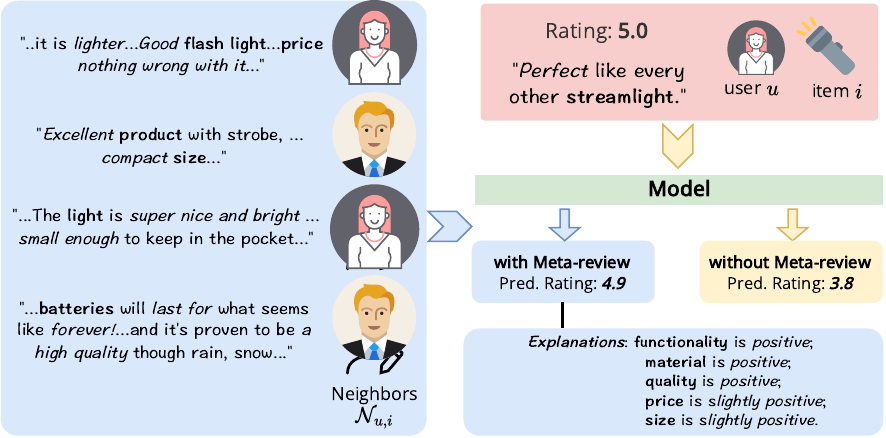}
    \caption{Case study to show the effectiveness of meta-review.}
    \label{fig:case}
\end{figure}

The model is designed to enhance the accuracy of rating prediction in RSs by leveraging reviews from neighbor users. Figure \ref{fig:case} presents a case to illustrate the effectiveness of incorporating these meta-reviews compared to relying solely on the target user's own review. When only using the user's own review, the limited information fails to comprehensively characterize the item, making it challenging for the model to accurately capture the user's overall preference on the target item with such a sparse contextual signal. Consequently, the model predicts a rating of 3.8, which deviates notably from the ground-truth rating of 5.0.

When leveraging the meta-review from neighbor users, this rich, aggregated information enables the generated meta-review to better model user preferences and complement the user's incomplete review, thereby explaining the perceived “perfect” evaluation. This approach effectively alleviates attribute incompleteness and review sparsity, leading to a much more accurate predicted rating of 4.9, closely matching the true score.

Furthermore, our model demonstrates interpretability through attribute-level sentiment prediction. The final explanation box aligns the overall positive review ("Perfect like every other streamlight") with specific positive sentiments on functionality, material, and quality, and mildly positive sentiments on price and size. This case highlights how meta-reviews successfully enhance both rating accuracy and explainability in real-world recommendation settings.
\section{Conclusion}
We present \textsc{Mosaic}, which addresses UGC \textbf{sparsity and incompleteness} by aggregating attribute-sentiment evidence from neighbor users. \textsc{Mosaic} 
combines an MMoE-based multitask architecture with a personalized attention module that conditions on the target user, jointly producing refined rating predictions and attribute-level explanations. Experiments on four real-world datasets show consistent gains over state-of-the-art baselines, and analyses confirm that meta-reviews expand attribute 
coverage beyond any single user's reviews and maintain consistent gains for users with limited interaction history.

\section*{Limitations}
We acknowledge several limitations that suggest directions for 
future research. First, \textsc{Mosaic} relies on a predefined domain-specific attribute taxonomy. Adapting to new domains requires reconstructing the attribute list, and emerging or niche attributes outside the predefined set cannot be captured. Future work could explore dynamic attribute 
discovery or open-vocabulary extraction. Second, neighbor 
selection in \textsc{Mosaic} is based on shared item interactions and 
optional social links, which may be sparse for highly isolated users. Integrating graph-based or behavior-similarity neighbor finding could further extend \textsc{Mosaic}'s applicability. Finally, \textsc{Mosaic} focuses on user-side neighbor-aggregated evidence. Integrating complementary item-side reasoning remains a promising direction for hybrid designs that further enhance performance in domains with rich item metadata.

\textsc{Mosaic} employs gpt-4.1-nano to extract attribute–sentiment pairs from existing user reviews, which raises potential concerns about LLM hallucination. We mitigate this risk through several design choices: the LLM is used only as an offline extraction tool and does not participate in training, inference, or the generation of any new content shown to users; its outputs are constrained to a predefined attribute taxonomy with a bounded sentiment vocabulary (limited to three words per opinion); and the extracted sentiments are further verified by a fine-tuned BERT classifier rather than being trusted directly. Most importantly, the meta-review construction aggregates evidence across multiple neighbor users via majority voting at the attribute level, which suppresses hallucinated extractions from any individual review. While these safeguards substantially reduce the impact of hallucination, we acknowledge that residual extraction errors may still propagate into the resulting meta-reviews and should be monitored in real-world deployments.

\section*{Acknowledgments}
This research is supported by the Collaborative Initiative, Interdisciplinary Graduate Programme (IGP), Nanyang Technological University, Singapore, and the Ministry of Education, Singapore, under its Academic Research Fund (AcRF) Tier 1 grant (RG16/25).

\bibliography{custom}

\newpage
\appendix
\section{LLM Prompt Template}
\label{app:prompt}

\begin{promptbox}
{\footnotesize
\textbf{Prompt for Attribute-sentiment Pair Extraction}

\smallskip
\raggedright
Analyze the review text.

\smallskip
\textbf{Requirements:}
\begin{enumerate}[leftmargin=1.5em, itemsep=2pt, topsep=2pt, parsep=0pt]
  \item Only consider these attributes: \texttt{\{ATTRIBUTES\_LIST\}}
  \item For each attribute, extract ONLY ONE opinion and limit it to a maximum of 3 words.
  \item If the attribute is not mentioned or sentiment is ambiguous, use \texttt{`N/A'}.
  \item The opinion must reflect CLEARLY POSITIVE or NEGATIVE sentiments.
\end{enumerate}

\smallskip
\texttt{\{REVIEW\_TEXT\}}

\smallskip
Output the extracted attribute-sentiment pairs in JSON format.
}
\end{promptbox}

\section{Experiment Details}
\subsection{Datasets}
\label{app:datasets}
All four datasets are publicly available for academic research under their respective terms of use, and the pretrained models (BERT and gpt-4.1-nano) are used in accordance with their license terms.
\begin{itemize}[leftmargin=12pt]
    \item \textbf{Yelp. \footnote{\url{https://www.yelp.com/dataset}}} This dataset comprises restaurant reviews collected from Yelp. It contains rich textual reviews and user-item interactions, making it suitable for modeling fine-grained review-based recommendations.
    \item \textbf{TripAdvisor. \footnote{\url{https://www.tripadvisor.com}}} This dataset includes user reviews on hotels from TripAdvisor. It provides detailed feedback and multi-aspect opinions, which are valuable for explainable recommendations.
    \item \textbf{Amazon Beauty and Amazon Sports. \footnote{\url{https://amazon-reviews-2023.github.io/}}} These two dataset consists of reviews from Amazon. Compared to the above two datasets, they represent a product-centric domain with highly subjective user experiences.
\end{itemize}

After preprocessing, each of the \textit{Yelp, TripAdvisor, and Amazon Sports} datasets retains approximately 100,000 high-quality user-item interactions with corresponding attribute-sentiment pairs. 

\subsection{Baseline}
\label{app:baselines}
\begin{itemize}[leftmargin=12pt]
    \item \textbf{PMF}~\cite{NIPS2007_d7322ed7} and \textbf{SVD++}~\cite{10.1145/1401890.1401944}: Classical matrix factorization models that estimate ratings by latent factors of users and items.
    
    \item \textbf{RGCL}~\cite{shuai2022review}: A relation-aware graph contrastive learning framework that constructs opinion-aware user-item graphs to capture semantic dependencies between opinions and interactions.
    
    \item \textbf{APH}~\cite{Liu_2025}: It implements review-based recommendations by jointly considering user preferences and item performance on specific aspects with a hypergraph. 
 
    \item \textbf{NETE}~\cite{li2020towards}: It models interactions between user/item embeddings and opinion words to generate fine-grained explanations.
    
    \item \textbf{PETER}~\cite{li2021personalized}: A pre-trained transformer-based model for explainable recommendation via a sequence prediction task.
    \item \textbf{PEPLER}~\cite{li2023personalized}: A prompt-enhanced pre-training model for personalized explanation generation.   
    \item \textbf{CER}~\cite{raczynski2023problem}: A transformer-based explainable recommendation framework focusing on coherent alignment between explanations and ratings. 
    \item \textbf{SERMON}~\cite{liao2025aspect}: A multi-modal recommendation framework based on contrastive learning and aspect-level explanations that enhances the interpretability.
    \item \textbf{LLMRec}~\cite{wei2024wsdm}: A recent LLM-based recommender that targets data sparsity by leveraging LLMs to augment the user-item interaction graph with generated edges and node attributes. For a fair comparison with \textsc{Mosaic}, we replace the original LLM backbone (gpt-3.5-turbo) with \textit{gpt-4.1-nano}, the same model used in our attribute-sentiment extraction pipeline.
\end{itemize}

\subsection{Experimental Settings}
\label{experimental_settings}
\textbf{Training Setting.} To make a fair comparison, each dataset is divided into training, validation, and test sets with a ratio of 8:1:1, where we adopt a per-user split and ensure at least one interaction is assigned to each set for every user. For the training stage, all models are optimized using the Adam optimizer, with the learning rate fixed at $1\times10^{-5}$. The embedding size for all models is set to 512. The training batch size is fixed at 512. For our model, the number of experts in the MMoE component is set to 3. All experiments are conducted on a single NVIDIA RTX A6000 48GB GPU and are repeated 5 times, with statistical significance analysis conducted to demonstrate the effectiveness of the proposed model.

\textbf{Evaluation Setting.} To assess rating prediction performance, we adopt the standard metrics of \textbf{Root Mean Squared Error (RMSE)} and \textbf{Mean Absolute Error (MAE)}. For explainability evaluation, our task focuses on attribute-sentiment prediction rather than generating textual explanations. Therefore, we adopt classical metrics from classification tasks— \textbf{Accuracy}, \textbf{F1 Score}, \textbf{Precision}, and \textbf{Recall}—to comprehensively evaluate the quality of predicted attribute-sentiment labels. For review generation baseline models, we derive the attribute-sentiment labels from textual explanations generated by each model, following the paradigm introduced in Section \ref{sec:llm_extraction}, and supervised by the user-level sentiment label $\mathbf{Y}_{u,i}$ for evaluation.

\subsection{Computational Cost Analysis}
\label{llm-analysis}

In this section, we analyze the computational efficiency of the proposed framework. As clarified earlier, the LLM doesn't directly add to our model's training or inference costs. Consequently, the computational cost reported in Table \ref{tab:cost} reports only the training-stage cost of the recommendation models and does not include the offline preprocessing.

As shown in Table \ref{tab:cost}, our model requires relatively low memory consumption and achieves fast training speed compared with representative baselines. The numbers in parentheses denote the number of training epochs determined by early stopping for each method. In particular, the per-epoch training time and total training time of our approach are comparable to lightweight models, and are substantially lower than those of generation-based or complex architectures. These results demonstrate that the proposed framework improves recommendation performance without introducing additional computational overhead, making it suitable for practical deployment.

\begin{table}[!t]
  \footnotesize
  \renewcommand{\arraystretch}{0.7}
  \setlength{\tabcolsep}{3pt}
  \centering
  \caption{Computational cost on the Sports dataset.}
  \label{tab:cost}
  \begin{tabular}{@{}lccc@{}}
    \toprule
    \textbf{Method} & \textbf{GPU Memory} & \textbf{\makecell{Train Time\\/Epoch}} & \textbf{Total Train Time} \\
    \midrule
    PETER & 1.78 GB & 9.3s & 1.5 min (10 epochs) \\
    PEPLER & 9.59 GB & 108.5s & 12.7 min (7 epochs) \\
    CER & 1.85 GB & 9.5s & 1.9 min (12 epochs) \\
    SERMON & 6.08 GB & 37.3s & 13.7 min (20 epochs) \\
    APH & 15.53 GB & 78.3s & 8.9 min (7 epochs) \\
    \midrule
    \textbf{Ours} & \textbf{1.57 GB} & \textbf{8.1s} & \textbf{1.3 min (10 epochs)} \\
    \bottomrule
  \end{tabular}
\end{table}

\subsection{Additional Analysis of Gate Distributions}
\label{gate_distri}

We emphasize that gating does not break down at larger $K$. Table ~\ref{tab:gate_k57} shows that at $K=5$ every attribute is still routed to a dominant expert and all five experts are used, yet RMSE degrades from 0.676 to 0.708 (+4.7\%). What changes monotonically is the sharpness of specialization: the average effective number of experts per attribute (exp of the gate entropy; 1.0 = a dedicated expert) rises from 1.51 ($K=3$) to 1.95 ($K=5$) to 2.39 ($K=7$). Additional experts do not create new specialists, they fragment existing ones instead. \textit{Rooms}, for instance, is routed almost entirely to one expert at $K=3$ but is split across two at $K=5$. Each expert then receives a weaker, more diffuse training signal while the parameter count grows, which is consistent with the reduced sample efficiency we observe. This explains that once experts are fragmented, the model pays the capacity cost of multiple experts without any of them becoming a specialist, which on TripAdvisor is enough to fall slightly below the single-expert setting (RMSE 0.708 / 0.713 vs. 0.696). Fragmentation is most severe at $K=7$, where the rating gate assigns near-zero weight to Expert 3 even though the attribute gates rely on it heavily, and only 4 of 7 experts are dominant for any attribute.

\begin{table*}[t]
\centering
\small
\caption{Mean gate weights at $K{=}5$ and $K{=}7$ on TripAdvisor (13,027 test samples). \textbf{Bold} marks the dominant expert per row. At $K{=}7$, the rating gate assigns near-zero weight (\underline{0.008}) to one expert, indicating expert redundancy under over-parameterization.}
\label{tab:gate_k57}

\begin{minipage}[t]{0.46\textwidth}
\centering
\setlength{\tabcolsep}{5pt}
\subcaption{$K=5$}
\begin{tabular}{lccccc}
\toprule
Task / Gate & E1 & E2 & E3 & E4 & E5 \\
\midrule
Rating        & 0.181 & 0.169 & 0.214 & 0.215 & \textbf{0.221} \\
\midrule
Service       & 0.002 & \textbf{0.988} & 0.004 & 0.000 & 0.006 \\
Cleanliness   & 0.009 & 0.009 & 0.014 & \textbf{0.782} & 0.186 \\
Rooms         & 0.004 & 0.089 & \textbf{0.648} & 0.240 & 0.019 \\
Location      & \textbf{0.712} & 0.215 & 0.012 & 0.014 & 0.047 \\
Price         & 0.010 & 0.046 & 0.114 & 0.106 & \textbf{0.725} \\
Sleep Quality & 0.008 & \textbf{0.922} & 0.003 & 0.016 & 0.050 \\
\bottomrule
\end{tabular}
\end{minipage}
\hfill
\begin{minipage}[t]{0.52\textwidth}
\centering
\setlength{\tabcolsep}{4pt}
\subcaption{$K=7$}
\begin{tabular}{lccccccc}
\toprule
Task / Gate & E1 & E2 & E3 & E4 & E5 & E6 & E7 \\
\midrule
Rating        & 0.089 & 0.239 & \underline{0.008} & 0.034 & \textbf{0.255} & 0.175 & 0.199 \\
\midrule
Service       & \textbf{0.682} & 0.017 & 0.041 & 0.108 & 0.090 & 0.059 & 0.003 \\
Cleanliness   & 0.029 & 0.009 & \textbf{0.773} & 0.038 & 0.004 & 0.064 & 0.083 \\
Rooms         & 0.038 & 0.029 & 0.028 & 0.008 & 0.007 & \textbf{0.883} & 0.006 \\
Location      & 0.003 & 0.030 & \textbf{0.667} & 0.234 & 0.043 & 0.002 & 0.020 \\
Price         & 0.050 & 0.006 & \textbf{0.781} & 0.022 & 0.017 & 0.047 & 0.077 \\
Sleep Quality & 0.004 & 0.006 & 0.006 & \textbf{0.719} & 0.224 & 0.006 & 0.033 \\
\bottomrule
\end{tabular}
\end{minipage}
\end{table*}

\section{Detailed Related Work}
\label{related_work}
\subsection{Explainable Recommender Systems}
Explanations will increase trust among users towards RSs by providing the recommendation along with a textual explanation. Explanations from extraction-based methods \cite{wang2022graph,pugoy2021unsupervised} usually consist of incomplete phrases instead of natural language. With advances of natural language processing, generation-based models, most of which rely on recurrent neural networks, RNN, e.g. LSTM\cite{cho2014learningphraserepresentationsusing}. NRT\cite{Li_2017} is a strong generation-based model that utilizes an RNN architecture to jointly generate more human-like textual explanations and predict ratings.

PETER\cite{li2021personalized} is the first generation-based explainable model that incorporates Transformer architecture to perform explanation generation, and PEPLER \cite{li2023personalized}, built upon PETER, employs pre-trained GPT-2 to enhance language understanding and reduce hallucinations. SEQUER \cite{ariza2023towards}, also extended from PETER, includes the user-review interaction sequence as the context of the language model to generate better explanations. POD\cite{li2023bprompt} is a Transformer-based model inspired by P5\cite{shuai2022review}. It adopts a pre-trained multitask LLM\cite{raffel2020exploring} as its backbone and is fine-tuned to perform top-N recommendation, next-item recommendation, and explanation generation. With the development of LLMs, they have been increasingly employed as sentiment extraction or classification tools in the process of explanation generation\cite{zhou2024enhancing,Zhang_2024}. 

To mitigate the hallucination of generation-based models, recent studies have adopted hybrid architectures to improve explanation tasks. ERRA \cite{cheng2023explainable} is composed of a retrieval mechanism to extract relevant attributes from the corpus, followed by a PETER-inspired architecture that generates explanations. ExBERT \cite{10096389} leverages Sentence-BERT to rank a user's historical reviews and selects top-K to construct pseudo user and item profiles, which are then encoded by a multi-head self-attention encoder to guide the BERT-based explanation generator. By grounding generation on retrieved reviews, it produces more personalized and aspect-aware explanations.

Regardless of whether a model is extraction-based or generation-based, existing approaches ultimately depend on the contextual features of reviews. As a result, they are inevitably subject to issues of hallucination and incompleteness, which makes it difficult to generate explanations that comprehensively cover all relevant attributes due to the inherent limitations of text generation models.

\subsection{Review-based Recommendataion}
Traditional RSs learn features using latent factor models that take user IDs and item IDs as input \cite{5197422}. However, these static feature learning methods suffer from data sparsity and scalability issues. Review-based recommendation aims to enhance rating prediction performance by leveraging the fine-grained semantic and opinion information embedded in user-generated reviews. 

Early CNN-based methods leverage CNNs to extract textual features from reviews. ConvMF\cite{kim2016convolutional} integrates CNN-based textual features into probabilistic matrix factorization (PMF) to enhance the accuracy of rating prediction. TransNet\cite{catherine2017transnets} introduces an additional transform layer that maps user and item features to an approximation of target review representation, thereby mitigating the reliance on reviews at test time.

By integrating an attention mechanism, NARRE\cite{chen2018neural} identifies the most useful features from individual reviews, thereby improving both accuracy and interpretability. D-Attn\cite{seo2017interpretable} extends this idea by applying dual attention across and within reviews, while DAML\cite{liu2019daml} applies dual and mutual attention to model interactions between user reviews and rating features. Similarly, CARL\cite{wu2019context} also jointly leverages textual and numerical signals to improve rating prediction performance. 

GNNs can capture the high-order relationships and model complicated user-item interactions by leveraging both semantic and structural information.  AHOR\cite{wang2023learning} and APH\cite{Liu_2025} capture aspects from reviews to model users' fine-grained preferences and item representations through GNNs. RGCL\cite{shuai2022review} and RMCL\cite{yang2023based} utilize contrastive learning techniques to enhance user-item embeddings and interaction modeling through the integration of review data, while MAGCL\cite{wang2023multi} improves preference granularity by disentangling reviews into multiple semantic aspects for contrastive learning. MCCL\cite{wei2024multi} further improves the capability of leveraging semantic information in a self-supervised manner. 

PLLMPAR\cite{li2023prompt} extracts key aspects that reflect user preferences and item properties using a fine-tuned language model. This study paves the way for further exploration and development at the intersection of LLMs and personalized recommendation.


\end{document}